\documentclass[a4paper,12pt]{article}

\usepackage{graphicx}
\usepackage{bm}
\usepackage{amssymb}
\usepackage{color,graphics}
\begin{document}

\begin{center}

\textbf{A mechanism that could stop the acceleration process within a collisionless shock}\\

\bigskip

Antoine Bret$^{1,2}$,
Asaf Pe'er$^3$
\\
\end{center}

$^1$ ETSI Industriales, Universidad de Castilla-La Mancha, 13071 Ciudad Real, Spain\\

$^2$ Instituto de Investigaciones Energ\'{e}ticas y Aplicaciones Industriales, Campus Universitario de Ciudad Real,  13071 Ciudad Real, Spain\\

$^3$ Bar-Ilan University, Ramat Gan, 5290002, Israel\\

\newpage

\begin{abstract}
Collisionless shocks are complex nonlinear structures that are not yet fully understood. In particular, the interaction between these shocks and the particles they accelerate remains elusive. Based on an instability analysis that relates the shock width to the spectrum of the accelerated particle and the shock density ratio, we find that the acceleration process could come to an end when the fraction of accelerated upstream particles reaches about 30\%. Only unmagnetized shocks are considered.
\end{abstract}


\newpage


\section{Introduction}
Shock waves are fundamental phenomena in fluids and plasmas. And collisionless shock waves represent a special kind of shock wave. Whereas in a shock wave in a fluid, the mean free path is very small compared to the dimensions of the system, in a collisionless shock wave, the mean free path is very large. For example, \emph{in situ} measurements showed the width of the Earth bow shock in the solar wind is about 100 km, while the proton mean free path at this location is about the Sun-Earth distance \cite{Bale2003,Schwartz2011}. This type of shock waves can only exist in a plasma, since they are mediated by collective electromagnetic effects instead of binary collisions \cite{Sagdeev66,balogh2013physics}.

In a collisional medium, any excess energy given to one particle is quickly shared with the others through collisions, on a time scale of the collision frequency. In a collisionless plasma, on the other hand, it is possible to give energy to one particle without it being immediately shared with the others.

It was thus realized in the late 70s that collisionless shocks can accelerate particles with a non-thermal spectrum of the kind $p^{-a}$, where $p$ is the momentum of the accelerated particles and $a$ is the power index \cite{Axford1977,Blandford78,Bell1978a,Bell1978b}.

Since the spectrum of cosmic rays detected on Earth from space obeys a power law \cite{Cronin1999}, or more precisely a succession of power laws, collisionless shocks in astrophysical media, such as those found in supernova remnant \cite{DrurySNR1995}, are excellent candidates for explaining the origin of cosmic rays. As such, they have been the subject of theoretical, numerical and experimental studies for decades (see \cite{Marcowith2016} and references therein).

Producing this kind of shocks in the laboratory, and observing accelerated particles, currently requires installations like the National Ignition Facility \cite{Fiuza2020}. Numerical studies, on the other hand, can only probe a short part of the shock's life after its formation \cite{Keshet09,Groselj2024}. It is therefore important to pursue theoretical studies which, while they can only address simplified models of the real process, allow to explore scenarios that are currently beyond the reach of experiments or simulations.

While cosmic rays acceleration in collisionless shocks has been explored since Fermi’s times \cite{Fermi1949}, the total amount of energy that goes into  cosmic rays is still unclear. An open question of interest to both theoreticians and observers who search for cosmic rays from various objects.

The aim of this paper is to propose, on a theoretical basis, a mechanism that could stop the acceleration process within a collisionless shock, and set a limit to the maximum fraction of particles promoted to cosmic rays.

Sections \ref{sec:conex} \& \ref{sec:origin} introduce the mechanism under scrutiny, which relies on an instability analysis. Section \ref{sec:disper} then derives the dispersion equation of the instability. Is is solved in Section \ref{sec:solve}, with the consequences discussed in Sections \ref{sec:csq} and following.

\begin{figure}
\begin{center}
\includegraphics[width=\textwidth]{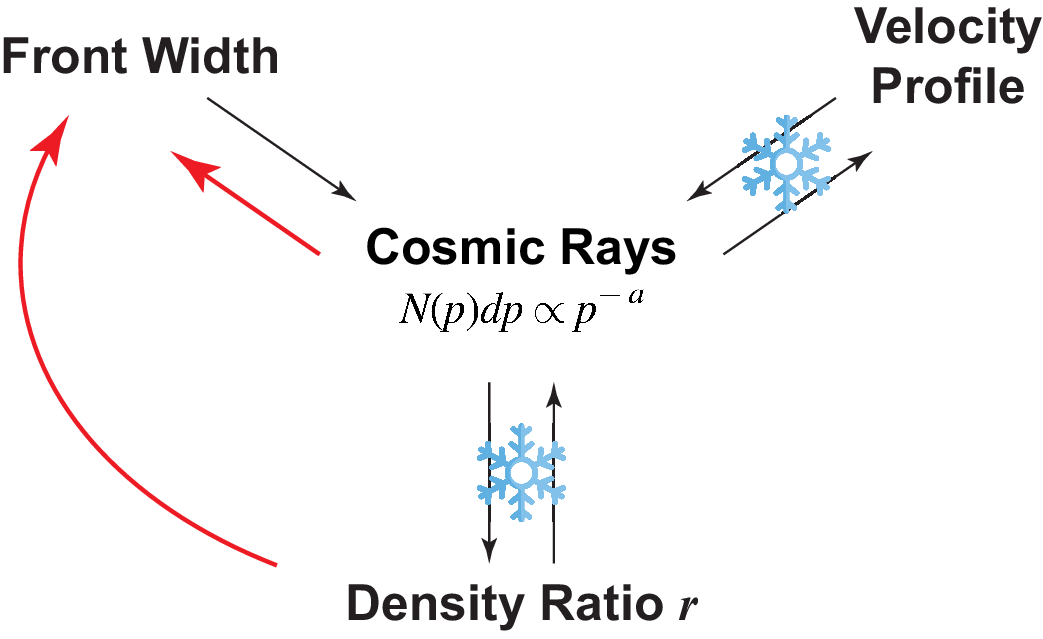}
\end{center}
\caption{The various ingredients of a collisionless shock and their connections. The red arrows represent the new connections proposed here. The snowflakes picture the connections we froze in the present work, namely, that we did not consider (see Section \ref{sec:disper}).} \label{fig:connections}
\end{figure}

\section{The ingredients of a collisionless shock and their connections}\label{sec:conex}
In order to introduce the mechanism we are proposing, we now present the various ingredients of a collisionless shock. They are schematically pictured on Figure \ref{fig:connections}.
\begin{enumerate}
  \item The density jump $r$ represents the ratio between the upstream and downstream densities. It is 4 in a strong sonic shock.
  \item The velocity profile represents the way the plasma velocity evolves spatially from the upstream to the downstream.
  \item The width of the shock front $\lambda$ represents the distance over which this transition takes place.
  \item Finally, cosmic rays represent the particles accelerated by the shock, mainly characterized by their power index $a$.
\end{enumerate}
As it happens, these 4 ingredients are interconnected. The power index $a$ depends on the density jump $r$ \cite{Blandford78}. But it also depends on the width of the shock front $\lambda$ \cite{DAS82}. For example, the greater the width of the shock front, the greater the power index. Furthermore, as we have just seen, a change in the power index $a$ is equivalent to a change in $r$, but the very presence of cosmic rays can also change $r$, when the energy they carry becomes substantial \cite{Eichler1984,Bret2020}.   Finally, there is a bidirectional relationship between $a$ and the velocity profile, since the latter can change the value of $a$ \cite{Schneider1987}, while cosmic rays, when their pressure becomes significant, in turn affect the velocity profile \cite{Blasi2002}.

We propose here some as yet unexplored links between these 4 ingredients, which could provide a mechanism to terminate the acceleration process. It is a causal relationship between the density jump $r$, the cosmic ray power index $a$ and the width of the shock front $\lambda$. These new connections are pictured by the red arrows on Figure \ref{fig:connections}.

\begin{figure}
\begin{center}
\includegraphics[width=\textwidth]{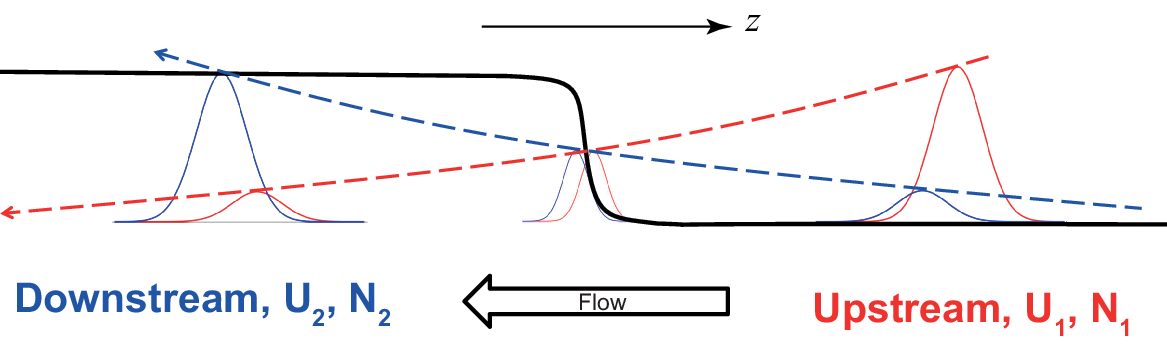}
\end{center}
\caption{The Mott-Smith \emph{ansatz}. The particle distribution function along the shock profile is a linear combination of the upstream and downstream  Maxwellians. The weight of the former vanishes as one progresses into the shock, while the weight of the latter grows. The far upstream and downstream flows have velocities and density $U_{1,2}$ and $N_{1,2}$ respectively.} \label{fig:mottsmith}
\end{figure}

\section{The origin of the new link proposed}\label{sec:origin}
The theoretical study of the density profile of a shock is a difficult subject. In 1951, Harold Mott-Smith proposed an \emph{ansatz} for studying such a profile within the framework of kinetic theory \cite{MS1951}. He hypothesized that the particle distribution function along the shock profile is a linear combination of the upstream and downstream  Maxwellians. Indeed this hypothesis has been validated by Particle-In-Cell simulations \cite{Spitkovsky2008}.

 In the Mott-Smith picture, see Figure \ref{fig:mottsmith}, the parameters of the upstream Maxwellian (density, drift speed, temperature) constitute the problem inputs; those of the downstream Maxwellian are given by the Rankine-Hugoniot relations, and only the respective weights of these 2 Maxwellians depend on the position $z$ along the shock profile. The weight of the upstream Maxwellian vanishes as one progresses into the shock, while the weight of the downstream one grows.

In 1967, Derek Tidman used Mott-Smith's model to study a collisionless shock \cite{Tidman67}. Among other results, he concluded that the thickness of the shock front can be given by an instability analysis at a location where the 2 Maxwellians have approximately the same weight. Since the upstream Maxwellian is centered around the upstream flow velocity $U_1$, while the downstream one is centered around the downstream flow velocity $U_2 = U_1/r$, this velocity shift results in an unstable system. Tidman analyzed the instability of this system in terms of the most unstable \emph{wavelength}, not the most unstable frequency, and related this wavelength to the thickness of the front, since it corresponds to the length over which the upstream flow is disrupted.

Now, Tidman's analysis did \emph{not} account for the presence of cosmic rays, that is, the particles accelerated by the shock. It only accounted for the unstable interaction between the two Maxwellians.

Here, we shall revisit Tidman's analysis including the population of cosmic rays accelerated by the shock. As we shall see, beyond a critical amount of upstream particles promoted to cosmic rays, the instability analysis provides a radically different answer to the question of the width of the front. In turn, such a change in the front width can radically change the power index of the cosmic rays.

Since Tidman's analysis was conducted for an unmagnetized shock, we only consider here the same kind of shocks, propagating without an external magnetic field.

\section{Dispersion equation}\label{sec:disper}
Figure \ref{fig:connections} makes it clear that it is not possible to elaborate a theoretical model accounting for all the connections involved in the problem. We therefore chose to ``freeze''  those signaled by the snowflakes on Figure \ref{fig:connections}. Hence, we set the density ratio $r$ to 4 in the sequel, ignoring the back reaction of the cosmic rays on the same ratio. We also considered a linear velocity profile between the upstream and the downstream, ignoring the back reaction of the cosmic rays on the same profile.

We therefore analyse the unstable system composed of the upstream and downstream Maxwellians, plus a population of cosmic rays. We implement a 1D model where location along the shock profile is identified by the $z$ coordinate. In order to simplify the phase space of parameters, we neglect temperature effects. The distribution function under scrutiny is therefore,
\begin{equation}\label{eq:df}
f(p) = n_1(z)\delta(p-m_iU_1) + n_2(z)\delta(p-m_iU_2) + n_{cr}f_{cr}(p),
\end{equation}
where $m_i$ is the ion mass, $\delta$ the Dirac delta function and $U_{1,2}$ the upstream and downstream flow speeds respectively. The first term refers to the upstream flow, the second one to the downstream flow, and the third one to the cosmic rays. They obey a power law spectrum of the form,
\begin{equation}\label{eq:fcr}
f_{cr}(p) = \kappa p^{-a}, ~~ p\in [P_{min} , +\infty],
\end{equation}
where
\begin{equation}\label{eq:kappa}
\kappa = \frac{a-1}{P_{min}^{1-a}},
\end{equation}
 is a normalization constant and $P_{min}$ the injection momentum.

We shall conduct the instability analysis at a location $z$ where $n_1 \sim n_2 = N_1/2$. Yet, we shall consider a fraction $\epsilon$ of the upstream flow has been ``promoted'' to cosmic rays so that we set $n_2 = N_1/2 - \epsilon$ and $n_{cr}=\epsilon N_1$.

The derivation of the dispersion equation is standard \cite{LandauKinetic} and yields,
\begin{eqnarray}\label{eq:disper}
\frac{4\pi \frac{N_1}{2}q^2 }{m_i}\frac{1}{(\omega - kU_1)^2} + \frac{4\pi (1-\epsilon) \frac{N_1}{2}q^2 }{m_i}\frac{1}{(\omega - kU_2)^2} \nonumber \\
 + \frac{4\pi \epsilon N_1 q^2 }{m_i} \kappa \int \frac{a p^{-1-a}}{\omega - kp/m_i}dp = 1,
\end{eqnarray}
where $q$ is the elementary charge. We now introduce,
\begin{eqnarray}
  \omega_{p1} &=& \frac{4\pi N_1 q^2}{m_i}, \\
   x &=& \frac{\omega}{\omega_{p1}}, \\
   Z &=&  \frac{kU_1}{\omega_{p1}}.
\end{eqnarray}
Considering $P_{min} \sim 2 m_i U_1$ \cite{Caprioli2014ApJ,Caprioli2015} and noting that $U_2 = U_1/r$, where $r$ is the compression ratio of the shock\footnote{$r$ will be set to 4 in Section \ref{sec:solve}, when solving the equation.}, Eq. (\ref{eq:disper}) eventually reads,
\begin{equation}\label{eq:disperOK}
\frac{1/2}{(x-Z)^2} + \frac{(1-\epsilon)/2}{(x-Z/r)^2} - \epsilon\frac{a(a-1)}{2xZ}\varphi\left(a,\frac{2Z}{x} \right) = 1,
\end{equation}
where,
\begin{equation}
\varphi(\alpha,\beta) = 2\int_1^{+\infty}\frac{t^{-\alpha-1}}{t^2\beta^2-1}dt.
\end{equation}
We now solve Eq. (\ref{eq:disperOK}) for $(Z,x) \in \mathbb{C} \times  \mathbb{R}$, since we are interested in the most unstable wavelength.

\begin{figure}
\begin{center}
\includegraphics[width=\textwidth]{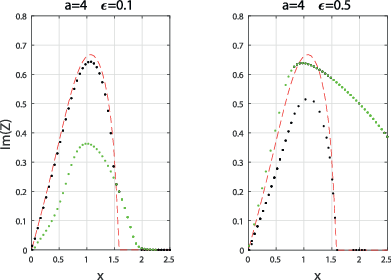}
\end{center}
\caption{Imaginary part of $Z$ solution of Eq (\ref{eq:disperOK}). Dashed red curve: two-flows unstable mode, that is, solution without cosmic rays, namely for $\epsilon=0$. Black circles: two-flows unstable mode, modified by the cosmic rays. Green circles: new unstable mode, triggered by the presence of the cosmic rays.} \label{fig:reso}
\end{figure}

\section{Resolution of the dispersion equation}\label{sec:solve}
Setting $r=4$, the numerical resolution of the dispersion equation reveals the presence of 1 unstable mode when cosmic rays are turned off, and 2 when they are turned on, as explained on Figure \ref{fig:reso}.

In the absence of cosmic rays, namely for $\epsilon = 0$ (red curves on Figure \ref{fig:reso}), there is only one unstable mode, arising from the interaction of the two flows. Its maximum spatial growth rate has Im$(Z) \sim 1$ and also Re$(Z) \sim 1$ (not shown). The corresponding most unstable wavelength has therefore,
\begin{equation}\label{eq:kmax}
\mathrm{Re}(Z) \sim 1 ~~ \Rightarrow ~~ \frac{1}{\mathrm{Re}(k)} \sim \frac{U_1}{\omega_{p1}}.
\end{equation}
Within Tidman's analysis, the width of the shock $\lambda$ is proportional to this quantity, up to a factor $A$, the ``Tidman's constant'', of order 10. We thus recover Tidman's result, established without cosmic rays, namely,
\begin{equation}\label{eq:lambda}
\lambda \sim A \frac{U_1}{\omega_{p1}}, ~~ A = \mathcal{O}(10).
\end{equation}
Suppose that instead of $\mathrm{Re}(Z) \sim 1$, we have $\mathrm{Re}(Z) \sim l$. Then,
\begin{equation}\label{eq:lambdaModif}
\lambda \sim A \frac{U_1}{\omega_{p1}}\frac{1}{l},
\end{equation}
which will be relevant shortly.

When cosmic rays are turned on, this two-flows unstable mode gets modified. It is pictured by the black circles on Figure \ref{fig:reso}. But now, another unstable mode appears (green circles), fully triggered by the presence of the cosmic rays. For small values of $\epsilon$, this new unstable mode grows slower than the two-flows one, and the width of the shock is not modified. But for higher values of  $\epsilon$, like  $\epsilon=0.5$ on Figure \ref{fig:reso}-right, this new unstable mode grows faster. It is therefore the real part of this most unstable $Z$ which now defines the width of the shock front.

\begin{figure}
\begin{center}
\includegraphics[width=\textwidth]{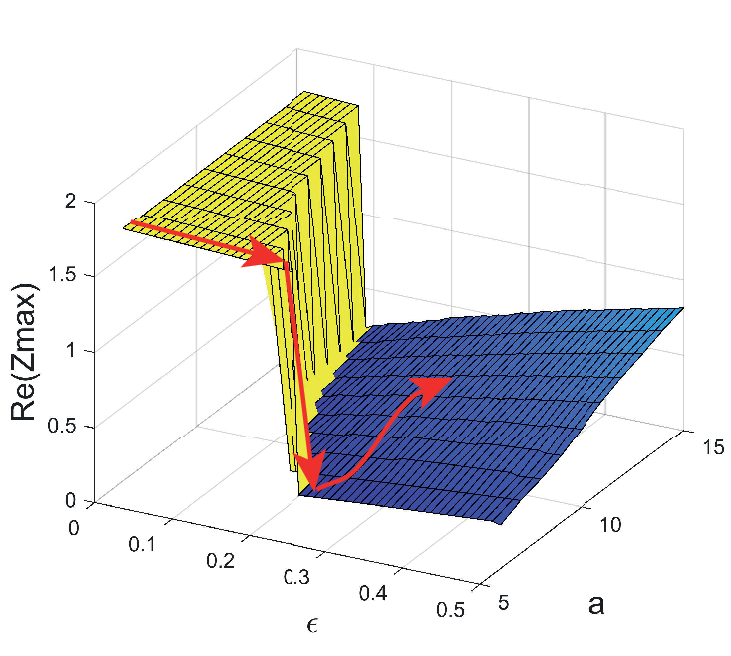}
\end{center}
\caption{Real part of the most unstable $Z$ solution of the  dispersion equation (\ref{eq:disperOK}). For small values of $\epsilon$, the most unstable mode remains the one triggered by the interaction of the upstream and downstream flows. But for $\epsilon > 0.3$, the most unstable modes arises from the presence of cosmic rays, with a Re$(Z)$ significantly smaller than before. The red arrows picture the temporal evolution of the system (see discussion in Section \ref{sec:discu}).} \label{fig:ReZ}
\end{figure}

We scanned the parameters phase space $(\epsilon, a) \in [0,1]\times [4,15]$. For each couple $(\epsilon, a)$, we computed the most unstable $Z$ and plotted its real part on Figure  \ref{fig:ReZ}.

One can navigate this plot following the time evolution of a strong collisionless shock. As it starts to propagate, $\epsilon \ll 1$ and $a \sim 4$, that is, the power index of a strong shock. The amount of accelerated particles then increases with time, but Re$(Z)$ does not, as evidenced by the plateau at low $\epsilon$.

Yet, once $\epsilon$ reaches $\sim 0.3$, an abrupt transition occurs. The most unstable mode switches from the two-flows one to the cosmic rays triggered one, and Re$(Z)$ suddenly falls from $1.65$ to $0.2$, almost 10 times smaller. As a result, and according to Eqs. (\ref{eq:kmax},\ref{eq:lambda},\ref{eq:lambdaModif}), the front width $\lambda$ increases almost 10 folds.

\section{Consequences on the power index $a$}\label{sec:csq}
Back in Section \ref{sec:conex}, we referred to the connection between the front width and the power index. In \cite{Blandford78}, the front was considered a discontinuity. Within this approximation, it was found that a shock could accelerate particles with a power index $a$ given by,
\begin{equation}\label{eq:aBO}
a = \frac{3r}{r-1},
\end{equation}
where $r$ is the density ratio. For a strong sonic shock with $r=4$, this gives $a=4$. It was then found in \cite{DAS82,Schneider1987} that considering a finite width $\lambda \neq 0$ for the front implies a larger value of $a$, because it is then more difficult for particles to go back and forth around the front and undergo Fermi cycles. Considering a simple linear transition over a distance $\lambda$ for the velocity field and $r=4$, the numerical result derived in \cite{Schneider1987} can be expressed as \cite{BretApJ2024},
\begin{equation}\label{eq:a(l)}
a \sim 4 + \frac{1}{6}\frac{\lambda}{D/U_1},
\end{equation}
where $D$ is the diffusion coefficient of the cosmic rays. Strictly speaking, this quantity depends on the location along the shock profile, and of the momentum of the particles. Here, like in \cite{Schneider1987,AS11}, we consider an average value of $D$.

The dimensionless quantity $\lambda U_1/D$ is the shock P\'{e}clet number. Equation (\ref{eq:a(l)}) is accurate up to $\lambda U_1/D \sim 15$, beyond which the exact value keeps growing, though slower than the linear trend \cite{BretApJ2024}.

\section{Discussion} \label{sec:discu}
We can now draw the conclusion of these calculations following the temporal evolution of the shock on Figure \ref{fig:ReZ}.

At the beginning of its history, the density jump is $r=4$, with a power index $a=4$. Then, as it propagates, $\epsilon$ grows, as the shock accelerates more and more particles. It then follows the trajectory indicated by the red arrow on the upper plateau.

When the fraction of accelerated particles reaches about 30\%, the most unstable mode becomes the one triggered by the cosmic rays, and the system falls onto the lower plateau, following the vertical red arrow downwards.

At this point, and according to Eqs. (\ref{eq:kmax},\ref{eq:lambda},\ref{eq:lambdaModif}), the width of the front increases. In turn, according to equation (\ref{eq:a(l)}), the power index increases. Suppose the front width $\lambda$ was such that before the transition we had $\lambda U_1/6D = 1$ in Eq. (\ref{eq:a(l)}), hence $a=5$. After the transition, with $\lambda$ increased 10 fold, $\lambda U_1/6D$ jumps from 1 to 10, and $a$ from 5 to 14. Such a high power index means acceleration stops, as the extension of the shock front becomes so large that it forbids Fermi cycles.

To our knowledge, the maximum fraction of upstream particles promoted to cosmic rays is an open question, mainly studied through PIC simulations. Some long ran simulations show values of $\epsilon$ reaching about 5\% \cite{Sironi2011,GS12,Caprioli2015}. Yet, the longest simulations to date only capture a short fraction of the shock lifetime \cite{Keshet09,Groselj2024}. Since the fraction $\epsilon$ grows with time \cite{Keshet09,Sironi2013,Caprioli2014ApJ}, it could reach the threshold commented here in real settings.

Strictly speaking, Eq. (\ref{eq:a(l)}), which relates the power index $a$ to the front width $\lambda$, assumes $\epsilon \ll 1$ since it is derived ignoring the back-reaction of the cosmic rays on the velocity profile\footnote{In \cite{Schneider1987} for example, up to 5 different velocity profiles were considered, without accounting for the back-reaction of the cosmic ray on them.}. Values of $\epsilon \sim 0.3$  invalidate this assumption. Therefore, as specified from Figure \ref{fig:connections} and in Section \ref{sec:disper},  we ignored, ``froze'', such back-reaction in order to obtain an analytically tractable model. Yet, we think the mechanism presented here could be robust enough to withstand a more realistic scenario.

Noteworthily, Tidman’s initial analysis was performed for an electrostatic shock whereas electromagnetic shocks mediated by the Weibel instability have a different structure \cite{Stockem2014}. However, as far as the shock width is concerned, the key ingredient here is the instability of the two, namely upstream and downstream, shifted Maxwellians. This coexistence of the two Maxwellians has also been observed in Particle-In-Cell simulations of electromagnetic shocks \cite{Spitkovsky2008}. We therefore think we can apply it for this kind of shocks.

Besides, the acceleration mechanism considered in our work is the  Fermi process contemplated in \cite{Blandford78,DAS82,Schneider1987} and for which Eq. (\ref{eq:a(l)}) stands. It differs from laser shock acceleration where the upstream ions gain energy from only one interaction with the shock \cite{SilvaPRL2004,Stockem2014,Boella2018}.

In summary, we have revisited Tidman's analysis of the width of a collisionless shock. While Tidman did not account for the shock accelerated cosmic rays, we did include them in our instability analysis. The result is that the most unstable wavelength is dramatically increased when the faction of accelerated particles reaches $\epsilon \sim 30\%$. Hence, we have uncovered an unexplored mechanism that could put an end to particle acceleration in a collisionless shock. In this picture, the width of the shock front is given by an instability analysis between the upstream and downstream flows. When this analysis takes into account the cosmic rays accelerated by the shock, the most unstable wavelength is radically altered when the fraction of accelerated particles exceeds around 30\%. This result is a potentially abrupt increase of the index of accelerated particles, which could simply mean the end of the acceleration process. Very long term Particle-In-Cell simulations could help validate our scheme.

\section{Data Availability}
The data used to support the findings of this study are
available from the corresponding author upon request.

\section{Conflicts of Interest}
The authors declare that they have no conflicts of interest

\section{Acknowledgments}
A.B. acknowledges the support from the Ministerio de Econom\'{\i}a y Competitividad of Spain (grant No. PID2021-125550OB-I00).
A.P. acknowledges the support from the European Research Council via ERC Consolidating Grant No. 773062 (acronym O.M.J.).
Thanks are due to Luis Silva for helpful inputs.

\bibliographystyle{apalike}
\bibliography{BibBret}

\end{document}